\documentclass[conference,compsoc]{IEEEtran}
\ifCLASSOPTIONcompsoc
  \usepackage[nocompress]{cite}
\else
  \usepackage{cite}
\fi
\ifCLASSINFOpdf
\else
\fi
\usepackage{tikz}

\usepackage{filecontents}
\usepackage{booktabs}
\usepackage{tabularx}
\usepackage{xspace}

\usepackage{enumitem}
\usepackage{listings}
\usepackage{color}
\usepackage{caption}
\usepackage{subcaption}
\usepackage{bm}
\usepackage{bbm}
\usepackage{tabularray}
\usepackage{multirow}
\usepackage{arydshln}
\usepackage{pifont}
\usepackage{enumitem}
\usepackage[export]{adjustbox}
\usepackage{threeparttable}
\UseTblrLibrary{diagbox}
\usepackage{algorithm}
\usepackage{algorithmic}
\usepackage{amsfonts} 
\usepackage{xurl}
\usepackage{amssymb}
\usepackage[most]{tcolorbox}
\usepackage{framed}
\usepackage{tikz}
\usepackage{comment}
\usepackage{fontawesome5}
\usepackage{array}
\usepackage[table]{xcolor}
\usepackage{colortbl}
\definecolor{HeaderColor}{RGB}{228,224,236}
\usepackage{pifont}
\usetikzlibrary{positioning,arrows.meta}
\usepackage[colorlinks=true,linkcolor=green!80!black,urlcolor=black,citecolor=black]{hyperref}

\definecolor{linkblue}{RGB}{5,20,200}

\definecolor{formalshade}{rgb}{0.95,0.95,0.97}
\definecolor{darkblue}{rgb}{0.14,0.22,0.52}
\newenvironment{takeaway}{

\MakeFramed{\advance\hsize-\width\FrameRestore}}
{\endMakeFramed}
\newcounter{takeaway}
\definecolor{darkgreen}{rgb}{0.0, 0.549, 0.0}
\definecolor{obsshade}{rgb}{0.95,0.97,0.95} 
\newenvironment{observation}{

\MakeFramed{\advance\hsize-\width\FrameRestore}}
{\endMakeFramed}
\newcounter{observation}
\newcommand{\moai}{\textsc{MoAI}\xspace}

\begin{document}
%
\title{SoK: Attack and Defense Landscape of Mobile On-device AI Systems}



%
\author{
\IEEEauthorblockN{
Yujin Huang\IEEEauthorrefmark{1},
Xin Zheng\IEEEauthorrefmark{2},
Xingliang Yuan\IEEEauthorrefmark{1},
Kwok-Yan Lam\IEEEauthorrefmark{4}
}
\IEEEauthorblockA{
\IEEEauthorrefmark{1}\textit{The University of Melbourne},
\IEEEauthorrefmark{2}\textit{RMIT University},
\IEEEauthorrefmark{4}\textit{Nanyang Technological University}
}
\IEEEauthorblockA{
\{jinx.huang, xingliang.yuan\}@unimelb.edu.au,
xin.zheng2@rmit.edu.au,
\IEEEauthorrefmark{3}kwokyan.lam@ntu.edu.sg
}
}




\maketitle

\begin{abstract}
Mobile on-device AI (\moai) systems that integrate locally deployed AI models with conventional mobile software components are emerging as a key paradigm for delivering intelligent functionality directly on end-user devices.
By moving inference from remote cloud services to the local mobile environment, such systems enable privacy-preserving, low-latency, and offline-capable AI functionality, yet introduce new security risks arising from the local storage of AI models.
This paper presents the first comprehensive
systematization of knowledge on \moai security, covering security pillars, attack landscape, and
defense landscape of \moai systems.
We further identify unresolved gaps in current attack and defense research and point to promising directions for future research in this emerging area.
Our work establishes the first systematic framework for understanding the attack and defense landscapes of \moai systems, serving as a foundation for building secure \moai systems and advancing research in this critical domain.
Companion resources are available at \href{https://github.com/Jinxhy/Awesome-MoAI-Security}
{\textcolor{linkblue}{\nolinkurl{https://github.com/Jinxhy/Awesome-MoAI-Security}}}.
\end{abstract}

%
\IEEEpeerreviewmaketitle

\section{Introduction}

\begin{figure}[!t]
    \centering
    \begin{subfigure}{\linewidth}
        \centering
        \includegraphics[width=\linewidth]{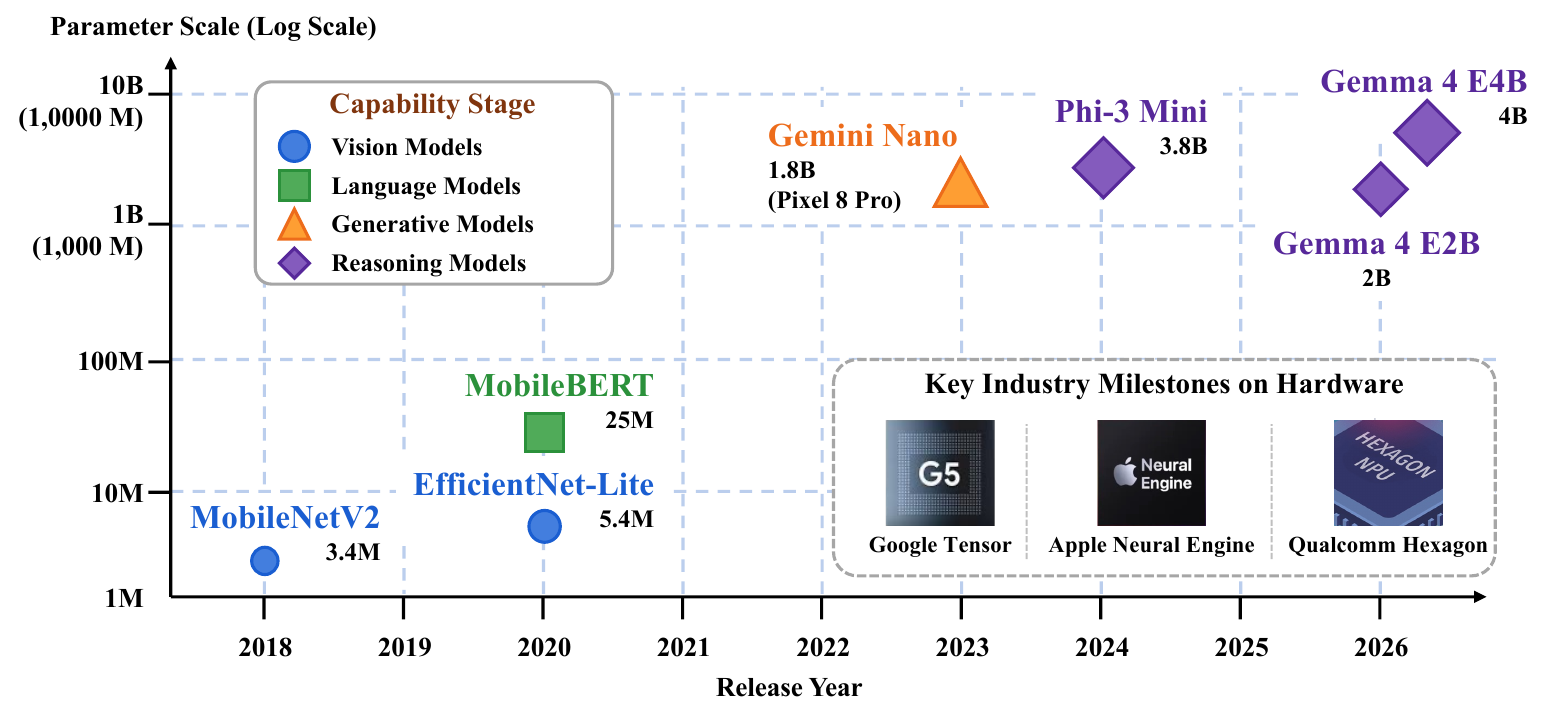}
        \vspace{-1.75em}
        \caption{Evolution of mobile on-device AI systems.}
        \label{fig:intro_overview_a}
    \end{subfigure}

    \vspace{0.7em}
    \begin{subfigure}{\linewidth}
        \centering
        \includegraphics[width=\linewidth]{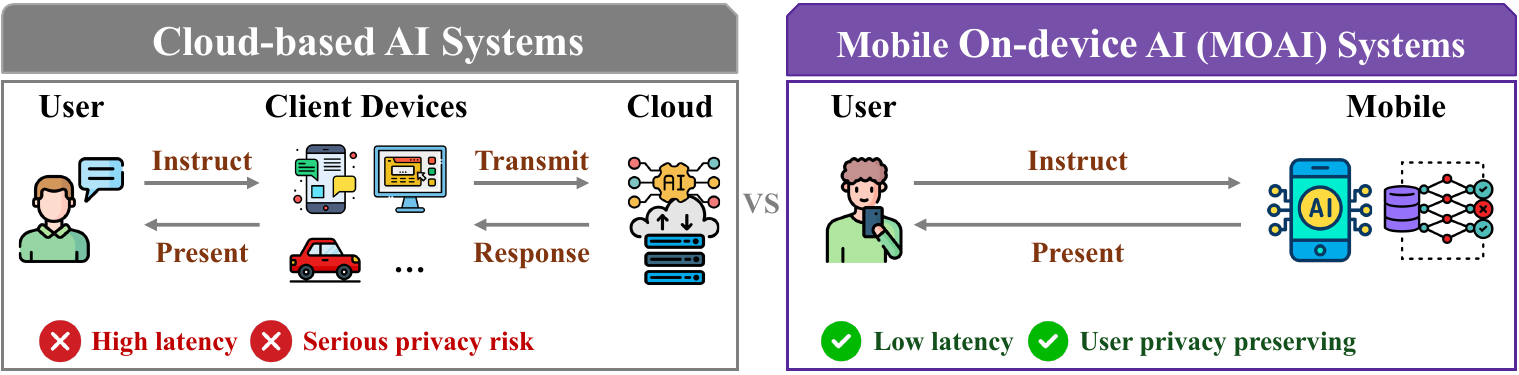}
        \caption{Comparison between cloud-based vs. \moai paradigms.}
        \label{fig:intro_overview_b}
    \end{subfigure}
    
    \vspace{0.7em}
    \begin{subfigure}{\linewidth}
        \centering
        \includegraphics[width=\linewidth]{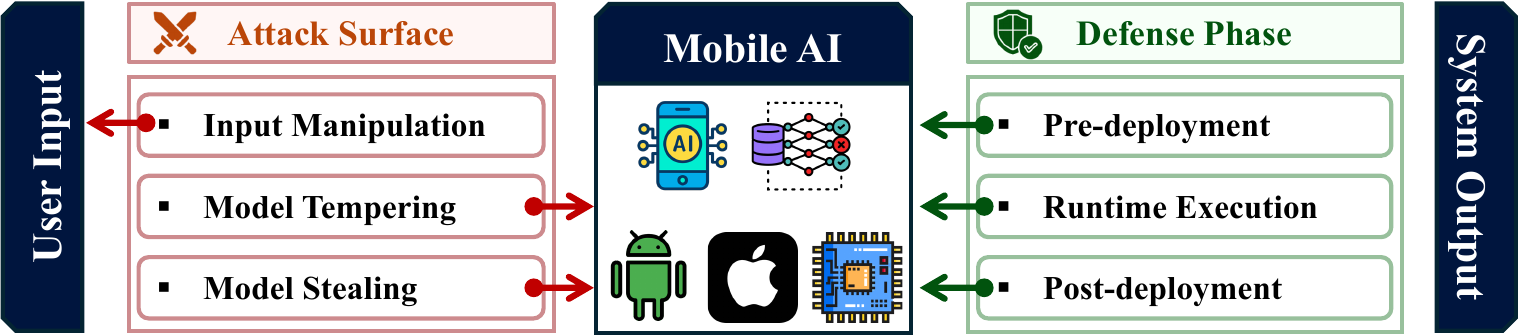}
        \caption{Attack surface and defense phase in \moai system.}
        \label{fig:intro_overview_c}
    \end{subfigure}
    \caption{Overview of the evolution, system paradigm, and security scope of \moai systems.}
    \label{fig:intro_overview}
    \vspace{-1.5em}
\end{figure}

\textbf{\underline{M}}obile \textbf{\underline{o}}n-device \textbf{\underline{AI}} (\moai) systems are emerging as a mainstream paradigm for delivering intelligent services directly on end-user smartphones.
Modern mobile apps increasingly rely on this paradigm to support AI-powered features such as image inpainting~\cite{li2024efficient}, extractive question answering~\cite{bohdal2025efficient}, and speaker diarization~\cite{broughton2023improving} without continuous cloud connectivity. 
As shown in Figure~\ref{fig:intro_overview_a}, this trend has been accelerated by the availability of dedicated AI hardware in latest mobile devices, such as Google Tensor~\cite{Google_tensor}, Apple Neural Engine~\cite{neuralengine}, and Qualcomm Hexagon~\cite{Qualcomm_hexagon}.
More recently, Google’s Gemma 4 family~\cite{gemma4}, particularly its E2B and E4B variants for mobile deployment, illustrates this shift as increasingly capable generative and reasoning models now target practical local deployment on end-user devices.
Compared to offloading AI from mobile devices to the cloud in Figure~\ref{fig:intro_overview_b}, \moai offers several distinct advantages.
These include improved privacy as sensitive data is processed locally on end-user devices, lower inference latency since there is no need to transmit data to remote servers, and continued functionality in scenarios where network connectivity is unavailable~\cite{xu2019first,huang2021robustness,sun2021mind}.

Despite the manifold benefits, \moai system inevitably stores models on end-user devices, which introduces new security threats.
As shown in Figure~\ref{fig:intro_overview_c}, an attacker can fully extract on-device models and inference-related information through techniques such as reverse engineering or device memory dumping.
This exposes multiple attack surfaces and facilitates a wide range of representative attacks, including adversarial attacks~\cite{huang2021robustness}, backdoor attacks~\cite{li2021deeppayload}, adversarial weight attacks~\cite{huang2026typhon}, model stealing attacks~\cite{sun2021mind}, and energy-latency attacks~\cite{wang2023energy}.
For instance, the attacker can reverse engineer a traffic sign recognition \moai system to obtain the on-device model, modify it to implant backdoors, reassemble it back to the system, and trigger the backdoors by placing carefully crafted triggers on specific traffic signs, posing risks to end-user safety.
In response to these threats, various defense mechanisms across different deployment phases have been proposed to protect on-device models, including model obfuscation~\cite{zhou2023modelobfuscator}, model authorization~\cite{hua2021mmguard}, model execution within trusted execution environments (TEEs)~\cite{bayerl2020offline}, and model watermarking~\cite{huang2025themis}.
For example, a mobile system developer can obfuscate an on-device model before deployment to hinder model extraction and prevent subsequent malicious exploitation.

Existing literature reviews primarily focus on the security of Edge AI and TinyML systems~\cite{wingarz2024sok,huckelberry2024tinyml}, and a recent survey on \moai systems solely concentrates on model stealing~\cite{nayan2024sok}.
\textit{The community therefore lacks a systematic view or framework for understanding the overall security risks of \moai systems as a whole, as well as guidance on the foundational security pillars and practical security solutions for \moai systems}.

This paper addresses this gap by \textit{providing the first comprehensive systematization of knowledge on the security landscape of \moai systems}.
We approach \moai security from a holistic mobile system perspective, studying how local inference reshapes the trust boundaries among user-governed inputs, device-resident model artifacts, mobile AI runtimes, and device-native execution environments. 
Our analysis first characterizes \moai systems as a layered architecture and derives three security pillars that capture their core protection properties.
We then systematize the attack landscape of \moai systems through a threat model analysis, an attack taxonomy organized around \moai threat models, and a cross-pillar analysis that reveals how the categorized attacks undermine the security pillars.
We further examine the defense landscape of \moai systems via security objectives, a taxonomy of existing defense mechanisms across deployment phases, and a cross-pillar assessment of how these protections cover, trade off, and fall short across the security pillars.
Through this systematization, we identify open problems across attack practicality and defense effectiveness, revealing that many \moai attacks rely on strong deployment assumptions, whereas existing defenses remain partial and incur nontrivial security, performance, and deployability tradeoffs.
Finally, we outline future directions for securing emerging \moai paradigms, including on-device training, on-device GenAI, and agentic \moai.
Our contributions are summarized as follows:
\begin{itemize}[leftmargin=*]
    \item \textbf{Security Pillars of \moai Systems.}
    We present a systematic framework that characterizes \moai system security via three security pillars, i.e.,  Pillar I: user-governed input integrity, Pillar II: device-resident model security, and Pillar III: device-native environment confinement. We further analyze how local input handoff, post-deployment model residency, and device-native execution reshape attack surfaces and defense requirements.

    \item \textbf{Comprehensive Attack Landscape and Taxonomy.}
    We develop a systematic attack taxonomy that characterizes \moai attacks through three threat models (i.e., input-level, model-level, and execution-level adversaries) and provide a comprehensive classification of five attack categories that include adversarial attacks, backdoor attacks, adversarial weight attacks, model stealing attacks, and energy-latency attacks, revealing how user-governed inputs, device-resident model artifacts, runtime materialization, and mobile hardware heterogeneity reshape attack practicability and impact.

    \item \textbf{Comprehensive Defense Landscape and Taxonomy.}
    We develop a systematic defense taxonomy that characterizes \moai defenses by deployment phase (i.e., pre-deployment, runtime execution, and post-deployment), and classify existing defenses into four categories that encompass model obfuscation, model authorization, TEE, and model watermarking, revealing how mobile deployment constraints affect protection across model artifacts, authorization paths, runtime states, and ownership.

    \item \textbf{Future Research Directions on \moai Systems.}
    We outline a forward-looking research agenda for \moai system security, including on-device training, on-device GenAI, and agentic \moai systems.
    
\end{itemize}
To the best of our knowledge, this is the first work to systematically analyze the security landscape of \moai systems. 
Our systematization provides a comprehensive framework for understanding \moai attack risks and defense strategies, guiding future research toward building secure \moai systems. 
This work can serve as a foundational reference for researchers and practitioners working with \moai systems.
\section{Overview}
\label{sec:overview}

\textbf{Scope.} We focus on security risks and defenses that are unique to, or significantly amplified in, \moai systems compared with traditional mobile software and mobile cloud-based AI systems. 
We first characterize \moai systems as holistic compositions of traditional mobile software components and local neural components, organized into four layers, including input interface, AI model artifact, runtime execution, and hardware isolation (\hyperref[sec:preliminary]{\S\ref*{sec:preliminary}}). 
We then analyze \moai-specific attacks by formalizing \moai threat models, categorizing attacks around them, and assessing the practicality of their deployment assumptions (\hyperref[sec:attack_landscape]{\S\ref*{sec:attack_landscape}}). 
We exclude general mobile app vulnerabilities and pure cloud-side AI attacks unless they directly interact with on-device inference. 
Next, we systematize \moai defenses across three deployment phases, with attention to their security-cost tradeoffs and limitations (\hyperref[sec:defense_landscape]{\S\ref*{sec:defense_landscape}}). Finally, we outlines future directions for emerging paradigms in \moai systems (\hyperref[sec:future]{\S\ref*{sec:future}}).

\textbf{Differences From Existing Surveys and SoKs}.
Existing surveys and SoKs on Edge AI and TinyML security examine adjacent but distinct deployment settings. 
Wingarz et al.~\cite{wingarz2024sok} study decentralized edge-intelligence deployments that span wearable and IoT sensors, embedded devices, and edge gateways, where AI computation is placed close to data sources rather than centralized in the cloud. 
Huckelberry et al.~\cite{huckelberry2024tinyml} focus on TinyML systems, where microcontrollers face severe memory, compute, and energy constraints together with physical exposure and side-channel risks. 
The closest work, Nayan et al.~\cite{nayan2024sok}, only considers on-device model confidentiality in \moai systems, without addressing the broader end-to-end security risks introduced by \moai system deployment.
In contrast, our work provides a unified systematization of knowledge of MOAI security from the holistic mobile-system perspective, deriving security pillars, systematizing attacks across adversarial capabilities, organizing defenses across deployment phases, and identifying open problems that capture how local mobile deployment reshapes \moai attack practicality and defense robustness.

\section{\moai Systems and Security Implications}
\label{sec:preliminary}

\begin{figure*}[t!]
\centering
\includegraphics[width=\linewidth]{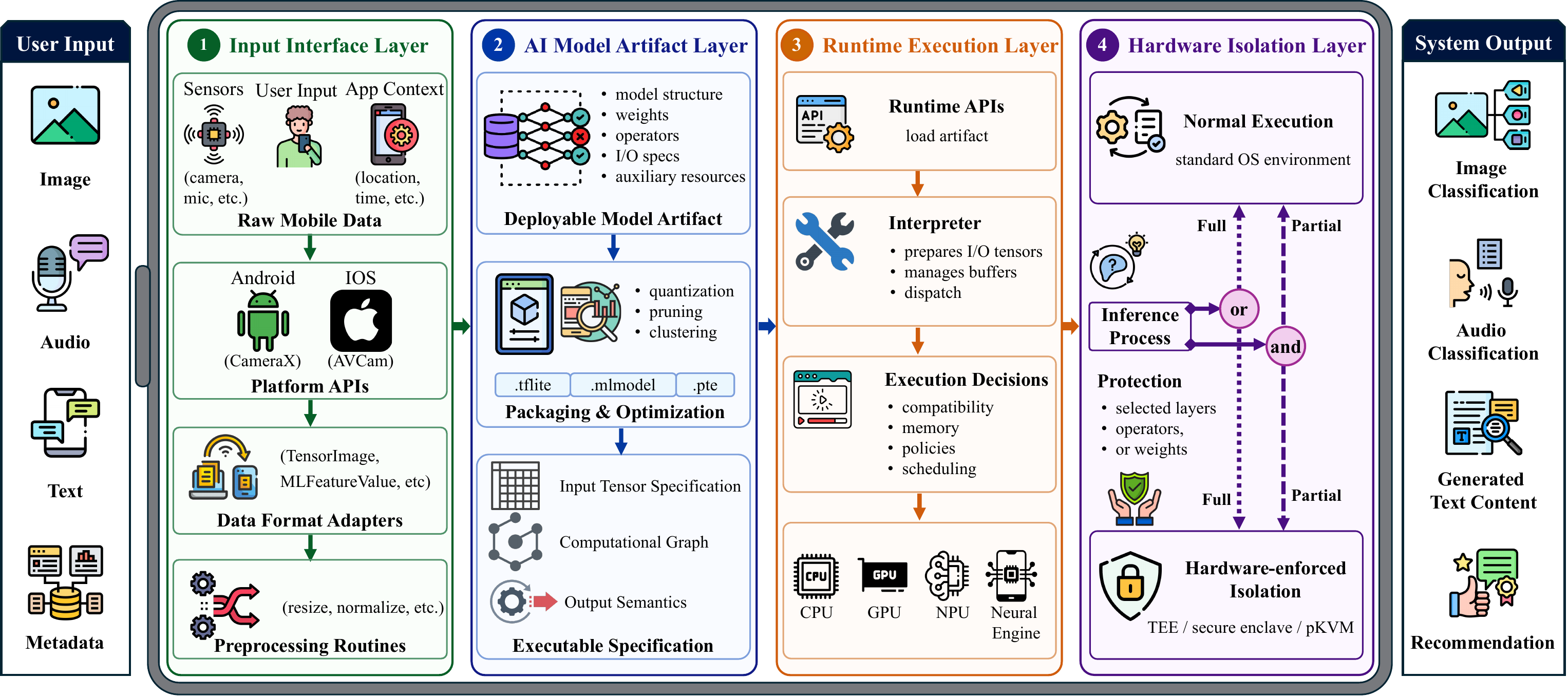}
\caption{Overview of a \moai system.}
\label{fig:system_overview}
\vspace{-1.5em}
\end{figure*}

\subsection{\moai Systems}
\label{sec:moai_systems}
In this paper, we define \moai systems as holistic compositions of traditional mobile software components and local neural components.
As shown in Figure~\ref{fig:system_overview}, a \moai system typically comprises the following layers.

\textbf{Input Interface Layer.}
The input interface layer defines the input boundary through which task-specific mobile data is acquired from sensors, user interactions and application contexts, and transformed into the input format required by the deployed on-device model.
It consists of device interfaces and input-handling code used before model invocation, including platform APIs such as Android CameraX~\cite{camerax} and iOS AVCam~\cite{avcam}, data format adapters such as LiteRT TensorImage~\cite{tensorimage} and CoreML MLFeatureValue~\cite{mlfeaturevalue}, and preprocessing routines such as image resizing and normalization.
Across deployments, this layer reconciles heterogeneous mobile input sources, platform-specific acquisition details, and modality-specific processing choices, then exposes them as input representations that conform to the deployed model’s input specification.
Consequently, downstream layers operate on structured model inputs rather than raw sensor buffers, input streams, or context dictionaries.

\textbf{AI Model Artifact Layer.}
As the core layer of a \moai system, the AI model artifact layer encapsulates trained neural models in deployable artifacts that can be stored on end-user devices, loaded by the local runtime, and invoked for inference.
These artifacts encode the model structure, trained weights, neural operators, input/output tensor specifications, and auxiliary resources such as labels, vocabularies, or tokenizers.
They are typically packaged into mobile-compatible formats such as LiteRT .tflite~\cite{litert}, CoreML .mlmodel~\cite{coreml}, or ExecuTorch .pte~\cite{executorch}, often with optimizations such as quantization, pruning, or clustering to meet constraints on storage, latency, and hardware compatibility.
After packaging, the artifacts expose concrete inference interfaces that specify input shapes, the computation graph to execute, and output semantics, and thus constitutes the executable specification for on-device inference.

\textbf{Runtime Execution Layer.}
The runtime execution layer translates executable model specification into concrete device-level computation under runtime, hardware, and resource constraints.
It loads AI model artifacts through local runtime APIs, initializes corresponding interpreters, prepares input and output tensors, manages intermediate buffers, and dispatches neural operators to available execution backends such as CPU kernels, GPU delegates, NPUs, or Neural Engine~\cite{neuralengine}.
These execution decisions are governed by operator compatibility, memory availability, backend selection policies, and scheduling constraints, which determine if computation is executed directly or delegated to accelerators.
Together, this runtime orchestration marks the final software-controlled stage before inference enters the hardware execution environment.

\textbf{Hardware Isolation Layer.}
Beyond runtime-managed execution, the hardware isolation layer characterizes the hardware-backed environment in which on-device inference is performed and the trust boundaries that govern access to inference code, model weights, and intermediate tensors.
It distinguishes the normal execution environment from hardware-isolated execution environments, such as trusted execution environments (TEEs), TrustZone, and protected kernel-based virtual machines (pKVMs).
Depending on the deployment design, inference may run entirely in the normal execution environment, be fully placed inside a hardware-isolated execution environment, or be partitioned so that only selected layers, operators, or weights are protected while the remaining computation proceeds through ordinary execution paths.
Ultimately, such isolation mechanisms serve as the hardware-enforced safeguard that protects on-device inference against software-level compromise.

\textbf{Overall Workflow.}
Given a user input, an \moai system first employs the input interface layer to transform it into the structured input representation required by the deployed model.
The AI model artifact layer then supplies the model artifact, whose executable specification determines the input tensor schema, computation graph, and output semantics for processing this structured representation.
Next, the runtime execution layer invokes the model artifact on the prepared input via available execution backends, while the hardware isolation layer governs if inference computation and state remain in normal execution or reside within isolated execution environments.
Finally, the system converts the model output according to the artifact-defined output semantics and returns it as a classification, generated content, or recommendation.
For instance, in an on-device skin cancer recognition system, a user-captured skin image is transformed into an input tensor, processed by a local vision model, and returned as a melanoma prediction with a dermatology referral.

\textbf{\moai Systems vs. Mobile Cloud-based AI Systems.}
While \moai and mobile cloud-based AI systems provide similar user-visible AI functionality, their key distinction lies in where the inference pipeline is executed.
In cloud-based AI systems, the mobile client primarily serves as the entry point to remote inference by acquiring task-specific data, constructing an inference request, and presenting the returned inference result. 
The cloud backend hosts the model artifact, executes inference, maintains intermediate inference state, and provides the required hardware resources.
In contrast, \moai systems move the inference pipeline onto end-user devices, where structured inputs, deployed model artifacts, runtime execution, and hardware isolation are composed within the local mobile stack.
This shift improves user privacy, inference latency, and offline availability, but also introduces new attack surfaces as models are stored locally.


\subsection{Security Pillars of \moai Systems}
\textbf{Pillar I: User-governed Input Integrity.}
This pillar characterizes the integrity of user-governed input consumed by \moai systems.
Here, user-governed input denotes the distinctive property of \moai systems that users ordinarily control input acquisition before inference.
Its integrity hinges on the trustworthiness of acquired input and the continuity from user-governed acquisition to model consumption.
A failure in either of these can make the system treat adversarial, unintended, or altered data as legitimate input, leading to unsafe behavior.
Preserving this integrity requires treating it as an end-to-end security property from user-governed acquisition to model-input handoff, so that data compromised at acquisition, unintentionally captured, or altered before inference cannot drive system behavior.

\textbf{Pillar II: Device-resident Model Security.}
This pillar characterizes the security of device-resident model artifacts in \moai systems.
Such proprietary artifacts are shipped to end-user devices for local execution, moving them outside the developer-controlled boundary.
This shift recasts model security as a post-deployment concern in which confidentiality shields deployed artifacts from unauthorized disclosure or extraction, and integrity safeguards them against malicious modification or replacement.
A confidentiality failure can expose model parameters and structures for unauthorized reuse, while an integrity failure can cause local inference to execute over attacker-modified model artifacts.
Securing device-resident models requires protection to encompass both the shipped artifacts and all post-deployment forms in which the models are stored, loaded, transformed, or materialized on devices.

\textbf{Pillar III: Device-native Environment Confinement.}
This pillar characterizes the confinement provided by the device-native environment in \moai systems.
The device-native environment encompasses the on-device software and hardware stack that mediates \moai inference, from the mobile OS and AI runtime to the memory subsystem and TEEs.
Its confinement captures the separation of \moai computation and sensitive runtime states from unintended observation and interference across the device stack, despite co-residence with untrusted system components and shared hardware backends.
When confinement fails, decrypted model buffers, runtime-resident inference states, and hardware-delegated memory regions may become visible and alterable to untrusted local components, rather than remain within the intended inference pipeline.
Effective confinement requires pipeline-level isolation, where on-device models remain protected throughout the OS, AI runtime, memory, and hardware stages that support inference.

\section{Attack Landscape of \moai Systems}
\label{sec:attack_landscape}

\begin{table*}[t]
\centering
\caption{Attack taxonomy for \moai systems.}
\label{tab:attack_taxonomy}
\small
\renewcommand{\arraystretch}{1}
\resizebox{\textwidth}{!}{
\begin{tabular}{p{4.5cm}p{4.5cm} p{7cm} p{3cm}} 
\toprule
\colorbox{red!50!gray!10}{\makebox[5cm][l]{\textbf{\faBiohazard \, Attack Category}}}
& \colorbox{red!50!gray!10}{\makebox[5cm][l]{\textbf{\faSitemap \, Sub-category}}}
& \colorbox{red!50!gray!10}{\makebox[7.5cm][l]{\textbf{\faBook \, Representative Works}}}
& \colorbox{red!50!gray!10}{\makebox[2.8cm][l]{\textbf{\faCrosshairs\, Threat Model}}}
\\

\midrule

\multirow{3}{*}[-2.75ex]{\textbf{Adversarial Attacks}}
& Model Similarity Exploitation
& \begin{tabular}[c]{@{}l@{}}Huang et al.~\cite{huang2021robustness}, Smart App Attack~\cite{huang2022smart}, \\ Deng et al.~\cite{deng2022understanding}, Cao et al.~\cite{cao2024cheating}, Hu et al.~\cite{hu2023first}
\end{tabular}
& \begin{tabular}[c]{@{}l@{}}Input-level \\Model-level\end{tabular} \\ \cmidrule{2-4}

&
Gradient Reconstruction
& \begin{tabular}[c]{@{}l@{}}REOM~\cite{zhou2024investigating}, 
TIM~\cite{wu2024tim}
\end{tabular}
& \begin{tabular}[c]{@{}l@{}}Input-level \\Model-level\end{tabular}\\ \cmidrule{2-4}

&
Preprocessing Manipulation
& Sang et al.~\cite{sang2023beyond}
& Input-level \\

\midrule

\multirow{3}{*}[-1.5ex]{\textbf{Backdoor Attacks}}
& Payload Injection
& DeepPayload~\cite{li2021deeppayload},
MalModel~\cite{hua2026malmodel}
& \begin{tabular}[c]{@{}l@{}} Input-level\\ Model-level\end{tabular} \\ \cmidrule{2-4}

&
Model Quantization
& Ma et al.~\cite{ma2023quantization}
& \begin{tabular}[c]{@{}l@{}} Input-level \\Model-level\end{tabular} \\ \cmidrule{2-4}

&
Image Steganography
& BARWM~\cite{wei2025stealthy}
& \begin{tabular}[c]{@{}l@{}}Input-level \\Model-level\end{tabular} \\

\midrule

\multirow{1}{*}{\textbf{Adversarial Weight Attacks}}
&
Parameter Tampering
&
TYPHON~\cite{huang2026typhon}
&
Model-level
\\

\midrule

\multirow{3}{*}[-2.75ex]{\textbf{Model Stealing Attacks}}


& Static Analysis
& 
\begin{tabular}[c]{@{}l@{}}Xu et al.~\cite{xu2019first}, Hu et al.~\cite{hu2023first}, \\ ModelXRay~\cite{sun2021mind}, REDLC~\cite{li2024redlc}
\end{tabular}
& Model-level \\ \cmidrule{2-4}

&
Dynamic Analysis
& \begin{tabular}[c]{@{}l@{}}ModelXtractor~\cite{sun2021mind};
AdvDroid~\cite{deng2022understanding};\\
DeMistify~\cite{ren2024demistify};
ArrowMatch~\cite{wang2025game}
\end{tabular}
& Execution-level \\ \cmidrule{2-4}

&
Side Channel
& Liu and Wang~\cite{liu2024model};
DeepCache~\cite{liu2024deepcache}
& Execution-level \\

\midrule

\multirow{1}{*}{\textbf{Energy-latency Attacks}}

&
-
&
Wang et al.~\cite{wang2023energy}
&
\begin{tabular}[c]{@{}l@{}}Input-level \\Model-level
\end{tabular}
\\

\bottomrule
\end{tabular}
}
\vspace{-1em}
\end{table*}

In this section, we systematize the attack landscape of \moai systems by defining threat models grounded in on-device adversary capabilities (\hyperref[sec:threat_models]{\S\ref*{sec:threat_models}}), developing a \moai-specific taxonomy for existing attacks (\hyperref[sec:attack_taxonomy]{\S\ref*{sec:attack_taxonomy}}), and examining how these attacks and their corresponding open problems map to the three \moai security pillars (\hyperref[sec:pillar_analysis_attack]{\S\ref*{sec:pillar_analysis_attack}}). Table~\ref{tab:attack_taxonomy} presents the attack taxonomy.

\subsection{Threat Models}
\label{sec:threat_models}

Given that \moai systems execute inference across user-governed inputs, device-resident model artifacts, and device-native execution environments, attackers and targeted assets vary across different scenarios. We identify three adversary types categorized by the attacker’s capabilities and points of adversarial manipulation in the on-device inference pipeline.

\textbf{Input-level Adversary.}
An input-level attacker is assumed to control the data supplied to an \moai system at inference time, while leaving the deployed model artifact and device-native execution environment unchanged~\cite{huang2021robustness,li2021deeppayload,deng2022understanding}.
The attacker aims to induce erroneous, targeted, or trigger-activated system behavior using inputs that appear benign or otherwise consistent with the intended task.
The attacker may possess task-level knowledge of the target system, such as input modality, task domain, and output semantics, and in stronger settings, model-level knowledge like trained weights and deployed model structure.
Successful attacks can shift classification, generation, or recommendation toward attacker-specified outcomes, creating user-facing safety and reliability failures without disrupting normal \moai operation.

\textbf{Model-level Adversary.}
A model-level attacker is assumed to access the model artifact that an \moai system stores, loads, or consumes on an end-user device, without relying on control over the inference-time input~\cite{xu2019first,sun2021mind,huang2026typhon}.
The attacker seeks to steal or compromise the deployed model by recovering its parameters and structure for unauthorized reuse or by altering its weights or computation graph to encode attacker-desired behavior.
Depending on the setting, the attacker may know regularities in mobile AI deployment, such as recognizable model file formats, framework APIs, operator sets, and model-loading interfaces.
Successful attacks can turn the deployed model into an extractable asset that can be copied and reused across various systems, or a compromised artifact that persistently steers local inference under benign inputs.

\textbf{Execution-level Adversary.}
An execution-level attacker is assumed to observe or interfere with the device-native execution process that loads model artifacts, initializes runtime states, and invokes \moai inference~\cite{sun2021mind,ren2024demistify,zhang2024no}.
The attacker intends to recover private inference data, confidential intermediate states, or protected models from runtime leakage exposed through memory materialization, partitioned execution, or side-channel observables.
Such recovery may rely on knowledge of execution-time state transitions, including model loading, memory allocation, TEE partitioning, accelerator offloading, and ciphertext observables.
Successful attacks can cause local execution to expose runtime state that should remain confined, compromising user privacy and model confidentiality without relying on malicious inputs or modified model artifacts.

\textbf{Practicality of Adversary Assumptions.}
These threat models differ not only in attacker capability, but also in deployment practicality. 
Input-level adversaries assume control over inference-time inputs, which is a strong and less practical assumption in \moai systems as inputs are normally governed by end users. 
Model-level adversaries are often more practical because \moai systems' model artifacts are shipped to end-user devices and can be analyzed, extracted, or modified from an attacker-controlled installation. Execution-level adversaries depend on where execution is observed. 
They are practical when attackers run \moai systems on their own instrumented devices, but become stronger when targeting another user’s live execution or hardware isolation. 
This distinction separates attacks enabled by ordinary post-deployment exposure from attacks that require control over victim-side inputs.

\subsection{Attack Taxonomy for \moai Systems}
\label{sec:attack_taxonomy}

As \moai systems directly run models on end-user devices, it introduces distinct attack surfaces across input, model, runtime, and hardware layers.
Based on the aforementioned threat models (\hyperref[sec:threat_models]{\S\ref*{sec:threat_models}}), we systematically examine existing \moai attacks and categorize them into five types: adversarial attacks, backdoor attacks, adversarial weight attacks, model stealing attacks, and energy-latency attacks.

\subsubsection{Adversarial Attacks}
\label{sec:adv_attacks}

Current studies on adversarial attacks against \moai systems primarily concentrate on how attackers leverage on-device model accessibility to craft adversarial examples that induce incorrect model outputs.
Based on the attack mechanisms, we categorize existing on-device adversarial attacks into three classes: \textit{model similarity attacks}, \textit{gradient reconstruction attacks}, and \textit{preprocessing manipulation attacks}.

\textbf{Model Similarity Exploitation.} This class of attack exploits the widespread reuse of AI models in mobile apps and constructs adversarial examples using surrogate models similar to target models. 
Early work \cite{huang2021robustness} introduced a pretrained model-based adversarial attack that compromises on-device models deployed in Android apps through identification of their highly similar pretrained models from TensorFlow Hub based on structure and parametric similarity.
Following this, Huang
et al.~\cite{huang2022smart} proposed Smart App Attack, a gray-box adversarial attack framework to hack on-device models by constructing highly
similar binary adversarial models based on identified transfer
learning approaches and pre-trained models, which successfully attacks 38 out of 53 real-world Android AI apps adopting transfer learning.

Subsequent studies breach the on-device model integrity under more constrained attacker capabilities.
Deng et al.~\cite{deng2022understanding} developed a semantic-based black-box adversarial attack that constructs substitute training datasets from Android AI app behavior to train surrogate models, enabling effective transfer-based adversarial example generation on on-device models.
Similarly, Cao et al.~\cite{cao2024cheating} proposed a practical black-box adversarial attack that instruments Android AI apps to collect query–response pairs, uses them to train substitute models, and transfers adversarial examples generated on the substitute ones to successfully fool original models.

In contrast to prior analyses of on-device models in Android ecosystems, Hu et al.~\cite{hu2023first} pioneered a security analysis of CoreML-based on-device models deployed in iOS AI apps by leveraging companion Android apps as attack surrogates to mount adversarial attacks, revealing practical attack surfaces introduced by cross-platform on-device model deployment.

\textbf{Gradient Reconstruction.}
Since on-device models are inference-only with backpropagation disabled, performing white-box adversarial attacks that outperform similarity-based gray-box or black-box attacks is challenging.
Motivated by this, Zhou et al.~\cite{zhou2024investigating} proposed REOM, an automated reverse-engineering framework that lifts inference-only constraints of on-device models by converting them back to backpropagation-supported versions via Open Neural Network Exchange~\cite{onnx}
interoperability, enabling effective gradient-based white-box adversarial attacks.
Along this direction, Wu et al.~\cite{wu2024tim} proposed
TIM, an automated white-box testing framework that reconstructs runnable IO processing code via precise Android static slicing and rebuilds inference-only on-device models into backpropagation-enabled counterparts by recovering computation graphs, stripped operator attributes, and parameter values for gradient-based adversarial attacks of \moai apps.

\textbf{Preprocessing Manipulation.}
Unlike the two aforementioned attacks, this class of attack targets the model preprocessing procedure from the model deployment perspective.
Sang et al.~\cite{sang2023beyond} first introduced a data
processing-based attack that injects malicious code snippets (e.g., manipulating image resizing, rotation, or RGB normalization) into the input preprocessing pipeline of real-world Android AI apps through repackaging, which affects model performance and inference latency without modifying on-device models.

\begin{takeaway}
\refstepcounter{takeaway}\label{op1}
\noindent\textbf{Open Problem \thetakeaway:}
\textit{The deployment of adversarial attacks against on-device models in \moai systems is non-trivial, as it necessitates input manipulation to apply adversarial perturbations or app repackaging to insert malicious code, which are impracticable and detectable once \moai systems are deployed on end-user devices.}
\end{takeaway}
\vspace{-1em}


\subsubsection{Backdoor Attacks}
\label{sec:backdoor_attacks}
Recent research has also explored implanting backdoors into on-device models in \moai systems, where attackers cause the models to exhibit predetermined malicious behaviors when processing inputs stamped with specific triggers.
According to the attack locus, we categorize existing on-device backdoor attacks into three
classes: \textit{payload injection attacks}, \textit{model quantization attacks}, and \textit{image steganography attacks}.

\textbf{Payload Injection.}
The core of these attacks lies in embedding backdoors into on-device models through data-flow graph modification, which mainly adds additional neural conditional branches into models as malicious payloads.
Li et al.~\cite{li2021deeppayload} introduced DeepPayload, the first payload injection–based backdoor attack, which disassembles an on-device model into a data-flow graph, adds a trigger detecting branch via graph manipulation, and recompiles the modified graph into a malicious model for backdoor activation.
Building on this work, Hua et al.~\cite{hua2026malmodel} proposed MalModel, which stealthily embeds malicious payloads within model parameters to produce malware Android AI apps.

\textbf{Model Quantization.} 
Quantization optimizes on-device models for efficient execution on resource-constrained mobile devices, yet creates opportunities for attackers to mount backdoor attacks.
Ma et al.~\cite{ma2023quantization} first explored quantization vulnerabilities in on-device models and proposed a post-training quantization backdoor attack that exploits truncation errors in int-8 weights during standard on-device quantization, awakening dormant backdoors in low-precision models while remaining undetectable in the full-precision ones.

\textbf{Image Steganography.} 
Another line of work investigates backdoor injection in on-device models without modifying the model structure to improve attack stealthiness.
Wei et al.~\cite{wei2025stealthy} presented BARWM, a stealthy backdoor attack against on-device models that utilizes image steganography to generate imperceptible and sample-specific triggers, then associates them with the models through backpropagation-supported model reconstruction and parameter reoptimization for backdoor injection.

\begin{takeaway}
\refstepcounter{takeaway}\label{op2}
\noindent\textbf{Open Problem \thetakeaway:}
\textit{Current backdoor attacks against on-device models in \moai systems must seek alternative entry points beyond standard training-phase insertion due to the read-only and inference-only nature of on-device models, which largely limits the development of stealthy attacks without causing observable changes.}
\end{takeaway}
\vspace{-1em}

\subsubsection{Adversarial Weight Attacks}
\label{sec:awv_attacks}
A more recent line of work has examined weight-level model tampering, where attackers directly modify the parameters of on-device models in \moai systems to induce incorrect or attacker-desired outputs.
Unlike adversarial backdoor attacks that rely on input manipulation, adversarial weight attacks compromise model behavior through parameter-level perturbation while preserving benign model utility.
As this attack category has only recently emerged, we do not categorize it and instead discuss the existing work.

Huang et al.~\cite{huang2026typhon} proposed TYPHON, the
first practical adversarial weight attack against on-device models in \moai systems. 
It circumvents the read-only and inference-only
constraints of mobile AI frameworks by reconstructing
writable model counterparts and computing malicious weights in a training-free manner. By rewriting selected weights, TYPHON can induce performance degradation or backdoor effects without sacrificing utility on benign inputs.

\begin{takeaway}
\refstepcounter{takeaway}\label{op3}
\noindent\textbf{Open Problem \thetakeaway:}
\textit{While adversarial weight attacks reveal a new parameter-level integrity failure in \moai systems, their deployment remains constrained by precise weight localization, as behavior-critical weights must be identified in a large and highly coupled parameter space without gradient guidance and under utility constraints.}
\end{takeaway}
\vspace{-1em}

\subsubsection{Model Stealing Attacks}
\label{sec:steal_attacks}
Beyond the aforementioned integrity-oriented attacks, model stealing attacks have received extensive attention as the local storage of on-device models naturally
enables attackers to disassemble \moai systems to obtain the models for intellectual property theft or downstream misuse.
On the basis of extraction strategies, we categorize on-device model stealing attacks into three classes: \textit{static analysis attacks}, \textit{dynamic analysis attacks}, and \textit{side-channel attacks}.

\textbf{Static Analysis.} 
This type of attack locates and extracts an on-device model via reverse engineering of \moai systems, combined with file scanning based on model naming schemes.
Xu et al.~\cite{xu2019first} first demonstrated the feasibility of stealing plaintext on-device models using Model Extractor, which scans the assets directory of decompiled mobile AI apps and identifies model files according to AI framework-specific suffixes, such as .tflite for LiteRT and .pb for TensorFlow.
Similarly, Hu et al.~\cite{hu2023first} leveraged Core ML model file suffixes (e.g., .proto and .mlmodel) to extract on-device models from iOS AI apps.
Meanwhile, Sun et al.~\cite{sun2021mind} designed ModelXRay to more precisely identify and extract on-device models by leveraging model suffix and AI framework–related magic words, such as LSTM, CNN, and RNN.
From a compiler-binary perspective, Li et
al.~\cite{li2024redlc} proposed REDLC, a learning-driven
reverse engineering pipeline that analyzes TVM-compiled
executables, recovers model computation graphs, infers
operator types and attributes from disassembled code, and
regenerates retrainable on-device models from compiled
binaries.

\textbf{Dynamic Analysis.}
Certain on-device models employ encryption to thwart static analysis attacks, but their runtime model decryption for inference in mobile AI apps exposes them to dynamic analysis attacks.
Sun et al.~\cite{sun2021mind} developed ModelXtractor that uses app instrumentation to dynamically locate memory buffers where decrypted on-device models are loaded and accessed by the corresponding AI frameworks, followed by in-memory model extraction.
Following this, Deng et al.~\cite{deng2022understanding} proposed AdvDroid, which hooks model-loading code, dynamically executes a mobile AI app to trigger model inference, and dumps models from memory after loading. 
Ren et al.~\cite{ren2024demistify} further advanced this dynamic analysis paradigm with DeMistify, an automated tool that combines static program slicing with runtime instrumentation to systematically identify, extract, and reuse on-device models and services from mobile AI apps at scale.
Recently, Wang et al.~\cite{wang2025game} proposed ArrowMatch, which breaks lightweight weight obfuscation
in TEE-shielded LLM partition schemes by matching direction distances between obfuscated offloaded weights and public pre-trained weights, thereby recovering private model weights.

\textbf{Side Channel.}
Complementing the instrumentation-based attacks, Liu and Wang~\cite{liu2024model} introduced a power side-channel based model extraction attack that profiles
open-source on-device models with a power monitor, trains a model architecture predictor from collected power traces, and uses runtime traces of victim inference to infer the model architecture without injecting malicious code into the target device.
At the cache level, Liu et al.~\cite{liu2024deepcache} proposed DeepCache, a cache side-channel based model extraction attack, which exploits cache-aware optimizations in compiler-generated model executables, collects Prime+Probe traces during inference, and infers operator types, hyperparameters, and optimized weights for model architecture stealing.

\begin{takeaway}
\refstepcounter{takeaway}\label{op4}
\noindent\textbf{Open Problem \thetakeaway:}
\textit{Despite local model storage in \moai systems on end-user devices, model stealing attacks remain far from trivial, because developers may employ customized encryption algorithms and AI frameworks, which substantially complicate reliable model identification, decryption, and extraction in practice.}
\end{takeaway}
\vspace{-1em}

\subsubsection{Energy-latency Attacks}
\label{sec:el_attacks}
In addition to compromising model integrity or confidentiality, existing research has also studied energy-latency attacks against on-device models, which aim to increase energy consumption and inference latency during \moai system execution. 
Different from the aforementioned attacks targeting prediction integrity and model confidentiality, this type of attack focuses on model availability under resource-constrained mobile deployment. Since such attacks remain largely underexplored in the \moai system, we discuss it separately rather than categorize it into a standalone taxonomy class.

Wang et al.~\cite{wang2023energy} first adapted sponge
poisoning to on-device models by poisoning training
examples to increase activation density and weaken
sparsity-based acceleration. Their on-device attack pipeline deploys poisoned models on mobile processors and simulates continuous inference over streaming inputs, showing that sponge poisoning can increase energy consumption and inference latency without sacrificing prediction accuracy.

\begin{figure*}[t!]
\centering
\includegraphics[width=\linewidth]{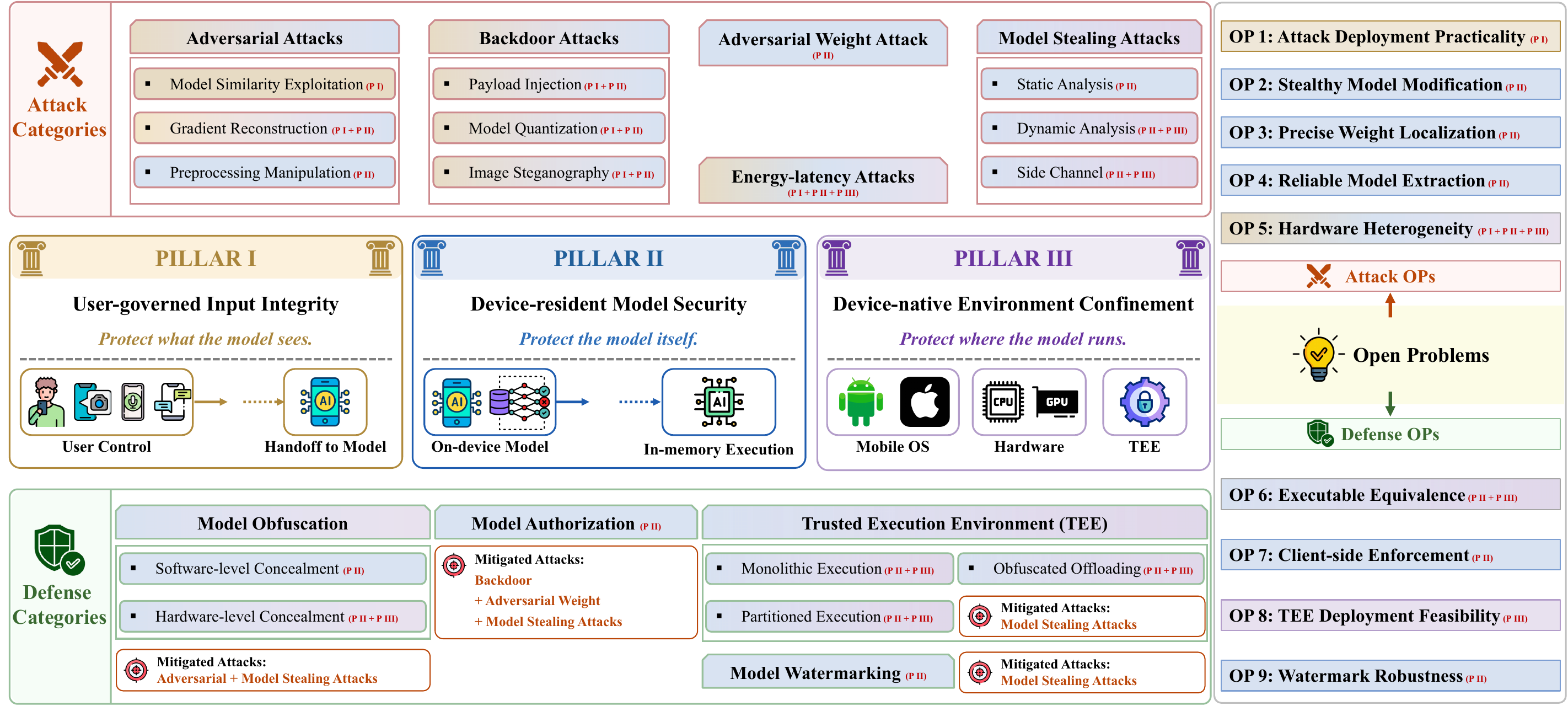}
\caption{Cross-pillar security analysis of attacks, defenses, and open problems in \moai systems.}
\label{fig:pillars}
\vspace{-1.5em}
\end{figure*}

\begin{takeaway}
\refstepcounter{takeaway}\label{op5}
\noindent\textbf{Open Problem \thetakeaway:}
\textit{The practicality of energy-latency attacks in \moai systems is constrained by mobile hardware heterogeneity, as poisoned activation patterns may amplify energy consumption and latency on sparsity-sensitive accelerators but remain ineffective on those lacking sparsity-dependent execution behavior.}
\end{takeaway}
\vspace{-1em}

\subsection{Cross-pillar Analysis of \moai Attacks}
\label{sec:pillar_analysis_attack}
Figure~\ref{fig:pillars} (the upper half) illustrates how the attack categories in \hyperref[sec:attack_taxonomy]{\S\ref*{sec:attack_taxonomy}} and their corresponding open problems (\hyperref[op1]{OP\ref*{op1}}-\hyperref[op5]{OP\ref*{op5}}) map to the three security pillars of \moai systems. 
Understanding these connections reveals that the security of \moai systems cannot rest on isolated safeguards for user-governed inputs, device-resident model artifacts, or device-native execution environments.
The mapping further shows that the identified OPs are shaped by deployment conditions that span pillars, since practical exploitation depends on coupled assumptions about input access, model exposure, and execution behavior.


\textbf{Security Pillars to \moai Attacks.}
The attack mapping reveals a security asymmetry in \moai systems, with Pillar II recurring as the central model-security concern, Pillar I remaining prominent in attacks that manipulate what the model consumes, and Pillar III surfacing only when attacks depend on hardware-mediated behavior.
Adversarial attacks primarily stress Pillar I as they manipulate the data presented to local inference, yet on-device model accessibility brings Pillar II into the attack path when surrogate models, reconstructed gradients, or deployment-specific preprocessing are used to strengthen the attacks.
The key deployment challenge for such attacks captured by \hyperref[op1]{OP\ref*{op1}} lies in delivering effective perturbations to model inputs through user-governed input paths.
Backdoor attacks couple Pillars I and II because trigger-stamped inputs only become harmful when the on-device model encodes the corresponding malicious behavior.
\hyperref[op2]{OP\ref*{op2}} therefore pinpoints the stealth problem in backdoor realization, where hidden malicious behavior needs to be introduced without causing observable changes to deployed model artifacts.
Adversarial weight attacks concentrate on Pillar II since they directly perturb deployed model parameters to alter inference behavior on targeted inputs.
At this parameter granularity, \hyperref[op3]{OP\ref*{op3}} centers on precision rather than access, as practical tampering depends on identifying behavior-critical weights in a large parameter search space while preserving benign utility.

Model stealing attacks shift from corrupting inference behavior to recovering the model artifacts protected by Pillar II, and extend to Pillar III when the recoverable forms of the model appear only after runtime loading, decryption, or execution.
Here, \hyperref[op4]{OP\ref*{op4}} separates model reachability from model recoverability, since local storage of on-device models does not ensure reconstruction of a complete, reusable model artifact in the presence of customized encryption schemes and nonstandard AI framework implementations.
Energy-latency attacks span Pillars I-III because availability degradation does not arise from inputs or models alone, but from their interaction with device-native scheduling, acceleration, and resource behavior.
Given this dependence on device-native execution, \hyperref[op5]{OP\ref*{op5}} captures the hardware-heterogeneity bottleneck, since the same cost-inflating behavior may not transfer across mobile stacks with different runtimes, accelerators, and resource-management policies.

\textbf{Takeaway.}
At the system level, \moai attack practicality turns on end-to-end deployment feasibility, governed by whether attackers can achieve input delivery, artifact modification, model recovery, or resource-cost amplification in the target mobile stack.
Such requirements arise because deployed \moai inference is not merely a direct model invocation.
User inputs are acquired through user-governed interfaces and transformed before model consumption, model artifacts are packaged, optimized, or encrypted before release, runtime states are materialized during loading and execution, and computation is scheduled across heterogeneous backends.
This deployment pipeline means that local model exposure does not directly translate into system-level exploitability, since access to on-device models does not ensure control over user-governed input paths, stealthy or precise artifact changes, complete model recovery, or predictable resource effects. 
\hyperref[op1]{OP\ref*{op1}}-\hyperref[op5]{OP\ref*{op5}} therefore characterize the practical gap between attacks enabled by local model access and attacks that remain feasible across the deployed \moai inference pipeline.



\section{Defense Landscape of \moai Systems}
\label{sec:defense_landscape}

In this section, we systematize the defense landscape of \moai systems by defining security objectives (\hyperref[sec:sec_ojb]{\S\ref*{sec:sec_ojb}}), developing a \moai-specific taxonomy for existing defenses across three deployment phases (\hyperref[sec:defense_taxonomy]{\S\ref*{sec:defense_taxonomy}}), and examining how these defenses and their corresponding open problems map to the three MOAI security pillars (\hyperref[sec:pillar_analysis_defense]{\S\ref*{sec:pillar_analysis_defense}}).
Table~\ref{tab:defense_taxonomy} presents the defense taxonomy.

\begin{table*}[t]
\centering
\caption{Defense taxonomy for \moai systems.}
\label{tab:defense_taxonomy}
\small
\renewcommand{\arraystretch}{1}
\resizebox{\textwidth}{!}{
\begin{tabular}{p{4.5cm}p{4.5cm} p{6.5cm} p{4.2cm}} 
\toprule
\colorbox{green!50!gray!10}{\makebox[5cm][l]{\textbf{\faShieldVirus \, Defense Category}}}
& \colorbox{green!50!gray!10}{\makebox[5cm][l]{\textbf{\faSitemap \, Sub-category}}}
& \colorbox{green!50!gray!10}{\makebox[7.5cm][l]{\textbf{\faBook \, Representative Works}}}
& \colorbox{green!50!gray!10}{\makebox[4cm][l]{\textbf{\faCrosshairs\, Mitigated Attacks}}}
\\
\midrule
\multirow{2}{*}[-1ex]{\textbf{Model Obfuscation}}

& Software-level Concealment
& \begin{tabular}[c]{@{}l@{}}ModelObfuscator~\cite{zhou2023modelobfuscator}, DynaMO~\cite{zhou2024dynamo}, \\ CustomDLCoder~\cite{zhou2024model}
\end{tabular}
& \begin{tabular}[c]{@{}l@{}} \hyperref[sec:adv_attacks]{\S\ref*{sec:adv_attacks}} Adversarial Attacks \\ \hyperref[sec:steal_attacks]{\S\ref*{sec:steal_attacks}} Model Stealing Attacks \end{tabular}\\ \cmidrule{2-4}

&
Hardware-level Concealment
& \begin{tabular}[c]{@{}l@{}}NNSplitter~\cite{zhou2023nnsplitter}, Mohseni et al.~\cite{mohseni2025novel}
\end{tabular}
& \hyperref[sec:steal_attacks]{\S\ref*{sec:steal_attacks}} Model Stealing Attacks \\

\midrule
\multirow{1}{*}{\textbf{Model Authorization}}

&
-
&
MMGuard~\cite{hua2021mmguard}
&
\begin{tabular}[c]{@{}l@{}} 
\hyperref[sec:backdoor_attacks]{\S\ref*{sec:backdoor_attacks}} Backdoor Attacks\\
\hyperref[sec:awv_attacks]{\S\ref*{sec:awv_attacks}} Adversarial Weight Attacks \\
\hyperref[sec:steal_attacks]{\S\ref*{sec:steal_attacks}} Model Stealing Attacks
\end{tabular}
\\
\midrule
\multirow{3}{*}[-6ex]{\textbf{\begin{tabular}[c]{@{}l@{}}Trusted Execution \\ Environment (TEE)\end{tabular}}}
& Monolithic Execution
& 
\begin{tabular}[c]{@{}l@{}}
Offline Model Guard~\cite{bayerl2020offline}, \\ GuardiaNN~\cite{choi2022guardiann}, Hu et al.~\cite{hu2023secure}, \\ T-Slices~\cite{islam2023confidential},
LEAP~\cite{sun2022leap},
ASGARD~\cite{moon2025asgard},\\
TZ-LLM~\cite{wang2026tz}, FlexServe~\cite{wu2026flexserve}
\end{tabular}
& \hyperref[sec:steal_attacks]{\S\ref*{sec:steal_attacks}} Model Stealing Attacks \\ \cmidrule{2-4}

&
Partitioned Execution
& \begin{tabular}[c]{@{}l@{}}
DarkneTZ~\cite{mo2020darknetz}, HybridTEE~\cite{gangal2020hybridtee},  \\ SecDeep~\cite{liu2021secdeep}, ShadowNet~\cite{sun2023shadownet},  \\ MirrorNet~\cite{liu2023mirrornet}, TSQP~\cite{sun2025tsqp},  \\ TEESlice~\cite{li2025teeslice}, TensorShield~\cite{sun2025tensorshield}
\end{tabular}
& \hyperref[sec:steal_attacks]{\S\ref*{sec:steal_attacks}} Model Stealing Attacks \\ \cmidrule{2-4}

&
Obfuscated Offloading
& 
\begin{tabular}[c]{@{}l@{}}
GroupCover~\cite{zhang2024groupcover}, ARROWCLOAK~\cite{wang2025game}, \\
ConvShatter~\cite{zheng2026miragenet}
\end{tabular}
& \hyperref[sec:steal_attacks]{\S\ref*{sec:steal_attacks}} Model Stealing Attacks \\

\midrule

\multirow{1}{*}{\textbf{Model Watermarking}}

&
-
&
THEMIS~\cite{huang2025themis}
&
\hyperref[sec:steal_attacks]{\S\ref*{sec:steal_attacks}} Model Stealing Attacks
\\
\bottomrule
\end{tabular}
}
\vspace{-1em}
\end{table*}

\subsection{Security Objectives}
\label{sec:sec_ojb}

In this section, we define the security objectives for \moai systems.
Such objectives are grounded in the standard \textit{Confidentiality}, \textit{Integrity}, and \textit{Availability} (CIA) triad.
In addition, we introduce \textit{Post-deployment Accountability} as an additional objective for capturing security requirements that arise after on-device models are released into uncontrolled deployment environments.

\textbf{Confidentiality.}
Confidentiality ensures that device-resident model artifacts and inference-related information in \moai systems are protected from unauthorized disclosure.
This protection covers model structures, trained parameters, computation graphs, operator types and attributes, input/output specifications, and runtime inference states.
This objective primarily addresses model stealing attacks (\hyperref[sec:adv_attacks]{\S\ref*{sec:steal_attacks}}), where adversaries extract on-device models through static analysis, dynamic analysis, or side-channel leakage, and further constrains downstream attacks that rely on access to model internals, including adversarial, backdoor, and adversarial weight attacks (\hyperref[sec:adv_attacks]{\S\ref*{sec:adv_attacks}},\hyperref[sec:backdoor_attacks]{\S\ref*{sec:backdoor_attacks}},\hyperref[sec:awv_attacks]{\S\ref*{sec:awv_attacks}}).

\textbf{Integrity.}
Integrity ensures that the inputs, model artifacts, and execution semantics of \moai systems remain resistant to unauthorized modification.
It includes preventing tampering with user-governed inputs, model preprocessing routines, model parameters, computation graphs, and inference outputs.
This objective addresses adversarial attacks (\hyperref[sec:adv_attacks]{\S\ref*{sec:adv_attacks}}), backdoor attacks (\hyperref[sec:backdoor_attacks]{\S\ref*{sec:backdoor_attacks}}), and adversarial weight attacks (\hyperref[sec:awv_attacks]{\S\ref*{sec:awv_attacks}}), where adversaries compromise local inference through malicious input perturbation, preprocessing manipulation, structural payload injection, quantization-induced model modification, or parameter-level tampering.

\textbf{Availability.}
Availability ensures that on-device inference in \moai systems is protected from adversarial degradation under resource-constrained mobile deployment.
This includes preserving acceptable inference latency, energy consumption, memory footprint, and execution stability across heterogeneous mobile runtimes and hardware backends. 
This objective addresses energy-latency attacks (\hyperref[sec:el_attacks]{\S\ref*{sec:el_attacks}}), where adversaries degrade inference quality by increasing computation cost or weakening hardware acceleration without necessarily affecting prediction accuracy.

\textbf{Post-deployment Accountability.}
As mentioned in \hyperref[sec:preliminary]{\S\ref*{sec:preliminary}}, on-device models are unique and essential to \moai systems because local inference requires these models to reside within the end-user mobile devices.
This deployment paradigm recasts model ownership as a post-deployment security concern, since on-device models may be extracted and reused after preventive protections fail.
Therefore, we introduce post-deployment accountability as a complementary security objective for \moai systems.
It ensures that the ownership of on-device models can be verified after suspected extraction or unauthorized reuse, providing the security basis for watermark-based ownership verification in the post-compromise phase.

\subsection{Defense Taxonomy for \moai Systems}
\label{sec:defense_taxonomy}

To mitigate the aforementioned attacks, various defense mechanisms have been proposed across different phases of the \moai system deployment lifecycle, including pre-deployment, runtime execution, and post-deployment, as illustrated in Figure~\ref{fig:intro_overview_c}.
In this section, we systematically examine \moai defenses that realize across these deployment phases and categorize existing defense mechanisms into three types: model obfuscation, model authorization, TEE, and model watermarking.

\subsubsection{Model Obfuscation}
Existing work on model obfuscation for \moai systems focuses on concealing on-device model structures, parameters, or execution dependencies to hinder model localization, white-box inspection, extraction, and unauthorized reuse after deployment. Based on the protection mechanisms, we categorize existing on-device model obfuscation defenses into two classes: \textit{software-level concealment} and \textit{hardware-level concealment}.

\textbf{Software-level Concealment.}
This class of defenses safeguards the binary representation of on-device models by obscuring the model during conversion, so as to restrict adversarial access to model structure and parameter information.
Zhou et al.~\cite{zhou2023modelobfuscator} first proposed ModelObfuscator, a prototype tool that obfuscates on-device models' structures and parameters through renaming, parameter encapsulation, neural structure obfuscation, shortcut injection, and extra layer injection to hinder static analysis and white-box adversarial attacks. 
Later, Zhou et al.~\cite{zhou2024dynamo} revealed that ModelObfuscator remains vulnerable to dynamic instrumentation and proposed DynaMO, which obfuscates on-device models by randomly coupling DL operators and applying linear weight transformations such that correct model information emerges only during paired execution.
To further obscure explicit model representations, Zhou et al.~\cite{zhou2024model} proposed CustomDLCoder, which replaces standard on-device model files with generated C/C++ implementations, thereby concealing model structures and parameters within native code to impede direct model localization and extraction.

\textbf{Hardware-level Concealment.}
Other studies safeguard on-device models at the hardware level by embedding access control and failure-inducing mechanisms into the model execution substrate, preventing unauthorized use even when models are exposed.
Zhou et al.~\cite{zhou2023nnsplitter} proposed
NNSplitter, which splits an on-device model into an exposed obfuscated submodel and a compact set of model secrets stored in trusted hardware. The exposed submodel alone produces degraded predictions, while authorized execution restores the original functionality by combining it with the protected secrets.
Subsequently, Mohseni et al.~\cite{mohseni2025novel} 
introduced a majority-logic-based method that obfuscates on-device models by transforming first-layer filters into a minority-derived weight configuration through hardware logic, thereby inducing unauthorized-use failure.

\begin{observation}
\noindent\textbf{Security-cost Trade-off:}
\textit{Model obfuscation improves confidentiality with generally modest latency cost, but its overhead shifts to memory, package size, native-code deployment, or hardware integration. For software-level concealment, ModelObfuscator~\cite{zhou2023modelobfuscator} lowers structural similarity to 0.67 with only an approximate 1\% latency overhead, but adds approximately 20\% RAM overhead and 3.2–59.8 MB package growth, DynaMO~\cite{zhou2024dynamo} reduces weight extraction from 98.76\% to 52.52\%, but adds 28.1\% RAM overhead, and CustomDLCoder~\cite{zhou2024model} removes explicit model artifacts to hinder model file parsing, but requires generated native-code deployment. For hardware-level concealment, NNSplitter~\cite{zhou2023nnsplitter} lowers unauthorized accuracy to 10\% by protecting only 0.002\% of weights, but incurs TEE-based secret storage and restoration requirements, and Mohseni et al.~\cite{mohseni2025novel} reduce wrong-key model accuracy to 1.0–9.5\% with reported 43\%, 79\%, and 71\% reductions in area, power, and weight-modification energy, but require specialized hardware integration.}
\end{observation}
\vspace{-1em}

\begin{takeaway}
\refstepcounter{takeaway}\label{op6}
\noindent\textbf{Open Problem \thetakeaway:}
\textit{The robustness of model obfuscation defenses in \moai systems remains bounded by the executable equivalence that obfuscation must preserve, as structures and weights hidden through transformed parameters, coupled operators, generated native code, or hardware-protected secrets still need to be composed into the original prediction function during authorized inference, exposing recoverable execution states for semantic, structural, and parameter recovery.}
\end{takeaway}
\vspace{-1em}

\subsubsection{Model Authorization}
Instead of concealing model internals, model authorization protects
on-device models by binding correct inference and model integrity to
an authenticated deployment context, so that unauthorized models lose
utility and modified models fail verification before inference.
As model authorization defenses for \moai systems remain underexplored, we discuss existing work in this line separately rather than organize it into finer-grained classes.

Hua et al.~\cite{hua2021mmguard} proposed MMGuard, the first automated framework for building mutual authentication between
Android AI apps and on-device models. It rewrites compiled models and mobile AI frameworks to add key-dependent input branches to selected layers and verify model hashes during initialization, so that packed weights can be correctly recovered only with the app-specific key generated from owner- and app-related signature information. As a result, stolen models lose prediction utility without the correct key, while tampered models fail verification and abort before inference, protecting against model stealing and tampering.

\begin{observation}
\noindent\textbf{Security-cost Trade-off:}
\textit{Model authorization improves confidentiality and integrity by making model utility depend on an authenticated app–model binding, but it incurs overhead from startup checks. MMGuard~\cite{hua2021mmguard} renders unauthorized or tampered models unusable, but its initialization overhead reaches 50–69\% (449 ms on average), and automatic protection succeeds on 37/43 evaluated mobile AI apps, with failures mainly caused by unsupported models or operators.}
\end{observation}
\vspace{-1em}

\begin{takeaway}
\refstepcounter{takeaway}\label{op7}
\noindent\textbf{Open Problem \thetakeaway: }
\textit{Model authorization in \moai systems protects on-device models by coupling correct inference with app-specific credentials and integrity verification, but the credential generation, model-hash verification, and packed-weight recovery must execute inside the mobile stack, making authorization enforcement dependent on client-side code that can be reverse engineered, repackaged, or instrumented after deployment.}
\end{takeaway}
\vspace{-1em}

\subsubsection{TEE}
Moving beyond artifact-level concealment or authorization, TEE defenses for \moai systems use hardware-isolated execution to protect on-device models during runtime, establishing trusted domains for full inference execution, security-sensitive model components, or obfuscated-output reconstruction.
According to the protected execution scope and computation placement, we categorize existing TEE defenses into three classes: \textit{monolithic execution}, \textit{partitioned execution}, and \textit{obfuscated offloading}.

\textbf{Monolithic Execution.}
The principle of these defenses is to execute the full on-device inference pipeline within TEE-backed runtimes, with the goal of safeguarding model parameters, intermediate data, and computation logic under mobile resource constraints.
Bayerl et al.~\cite{bayerl2020offline} proposed Offline
Model Guard, which leverages SANCTUARY~\cite{brasser2019sanctuary} to create user-space enclaves, attests to the enclave before encrypted model provisioning, and decrypts the model only inside the enclave to securely process peripheral inputs for on-device inference.
To address memory exposure and latency bottlenecks in such
secure execution, Choi et al.~\cite{choi2022guardiann} introduced
GuardiaNN, a TrustZone-based framework that keeps model data
encrypted in DRAM, confines plaintext computation to the secure
world, and uses direct convolution, SRAM-aware data reuse, and
cryptographic hardware to lower overhead.
Hu et al.~\cite{hu2023secure} further reduced secure memory pressure with a mobile TEE framework that combines progressive pruning, memory reclamation, and adaptive model loading to fit the entire inference process within constrained TEE memory.

Later studies shifted this direction toward more flexible TEE-backed inference under mobile deployment constraints.
Islam et al.~\cite{islam2023confidential} proposed T-Slices, a dynamic fragmentation framework, which divides unmodified model layers into memory-fit slices, sequentially loads them into TrustZone, and keeps inactive slices encrypted in untrusted memory to protect the entire inference execution.
At the application level, Sun et al.~\cite{sun2022leap} developed LEAP, a developer-friendly solution that extracts on-device models and inference codes from AI apps and runs them in lightweight isolated sandboxes with verified launch, exclusive peripheral access, and flexible resource control to shield sensitive inference from compromised rich-OS access and interference.
Beyond TrustZone-only designs, Moon et al.~\cite{moon2025asgard}
introduced ASGARD, a virtualization-based TEE framework that
extends protected execution to SoC-integrated accelerators through
secure I/O passthrough, reduces the trusted computing base via
platform and application debloating, and mitigates TEE-to-REE exits
with model execution planning.

For on-device LLM inference, Wang et al.~\cite{wang2026tz} proposed TZ-LLM, an Arm TrustZone-based secure inference system that runs the LLM framework as a trusted application, combines pipelined parameter restoration with dynamic secure-memory scaling, and uses a minimal TEE-side NPU co-driver to protect LLM parameters, activations, and KV cache against compromised REE and DMA-based extraction.
Similarly, Wu et al.~\cite{wu2026flexserve} presented FlexServe, which supports TrustZone-based mobile LLM serving with page-granular secure memory and switchable NPU execution to protect model weights, KV caches, and intermediate results from compromised-kernel access.

\textbf{Partitioned Execution.}
Given limited TEE memory and costly accelerator access, partitioned execution has emerged to place security-sensitive model components or framework routines inside the TEE and delegate less sensitive workloads to the untrusted normal world.
Mo et al.~\cite{mo2020darknetz} proposed DarkneTZ, a layer-level
partitioning framework that profiles a model's layer privacy sensitivity and executes sensitive layers inside the TEE while leaving the remaining layers in the normal world to reduce privacy risks with limited overhead. 
Similarly, Gangal et al.~\cite{gangal2020hybridtee} introduced HybridTEE, which bridges local ARM TrustZone and remote Intel SGX through a trusted channel and assigns model partitions across the two
TEEs to alleviate local secure-memory constraints. 
Then, Liu et al.~\cite{liu2021secdeep} presented SecDeep, a TrustZone-based framework with a split ARM NN design to keep plaintext tensor computation inside the TEE and protect normal world preparation and accelerator configuration through integrity checks and encryption. 
Later, Sun et al.~\cite{sun2023shadownet} proposed ShadowNet, which transforms linear-layer weights before untrusted accelerator offloading, keeps nonlinear operations inside the TEE, and restores outsourced results in the TEE to preserve model confidentiality.

Recent works reduce the trusted footprint from coarse execution
partitions to compact trusted logic and critical model states.
Liu et al.~\cite{liu2023mirrornet} proposed MirrorNet, a TEE-friendly framework in which a degraded BackboneNet executes in the normal world and a lightweight Companion Partial Monitor resides in the secure world to rectify outputs for authorized inference.
Subsequently, Sun et al.~\cite{sun2025tsqp} introduced TSQP, a quantization model partition framework that shields crucial re-quantization scales inside the TEE, offloads 8-bit operations to the normal environment, and uses parameter de-similarity with an integrity monitor to preserve model confidentiality and integrity.
Additionally, Li et al.~\cite{li2025teeslice} designed TEESlice, which partitions models before training into a public backbone and privacy-sensitive slices, prunes these slices to fit TEE memory, and offloads backbone linear layers to GPUs over encrypted and verifiable channels.
Furthermore, Sun et al.~\cite{sun2025tensorshield} presented TensorShield, a tensor-level partitioning framework that uses attention-transition analysis to identify critical tensors, protects selected tensors through TEE shielding, and combines critical-feature identification with latency-aware placement to maintain model confidentiality.


\textbf{Obfuscated Offloading.}
Since heavy neural operators are often delegated to untrusted accelerators for efficiency, this class protects outsourced inference through the exposure of only transformed weights and TEE-based recovery of authentic outputs.
Zhang et al.~\cite{zhang2024groupcover} proposed GroupCover, which applies reverse clustering and mutual covering to obfuscate convolution kernels before GPU offloading, and reconstructs authentic layer outputs inside the TEE to resist model stealing.
At the LLM scale, Wang et al.~\cite{wang2025game} introduced ARROWCLOAK, a lightweight obfuscation algorithm that randomizes private weight-vector directions inside the TEE before GPU offloading to mitigate direction-similarity-based weight recovery, with efficient TEE-side reconstruction of original layer outputs.
Recently, Zheng et al.~\cite{zheng2026miragenet} designed ConvShatter, a convolution obfuscation scheme that decomposes kernels into shared patch bases and kernel-specific damaged components, injects decoy kernels, permutes channels and kernels, and uses TEE-sealed metadata to recover GPU-computed obfuscated outputs.


\begin{observation}
\noindent\textbf{Security-cost Trade-off:}
\textit{TEE defenses improve model confidentiality through hardware-backed isolation, but their overhead shifts to secure-memory pressure, TEE/REE coordination, and accelerator integration. For monolithic execution, TZ-LLM~\cite{wang2026tz} protects on-device model parameters and runtime states with TrustZone/NPU support, but incurs 5.2–28.3\% TTFT overhead and 1.3–4.9\% decoding overhead over REE inference. For partitioned execution, TensorShield~\cite{sun2025tensorshield} achieves near full-TEE protection with up to 25.35× speedup over prior TEE-based inference, but incurs offline profiling and selection costs, including 23 min–1 h 25 min hardware profiling and 10.9 average evaluation epochs for critical-tensor selection. For obfuscated offloading, ARROWCLOAK~\cite{wang2025game} increases weight-direction distance by over 900× and improves defense by 6.5×, but adds 0.46× overhead over non-obfuscated TSLP, with TEE-side recovery taking 40.70\% of runtime.}
\end{observation}
\vspace{-1em}

\begin{takeaway}
\refstepcounter{takeaway}\label{op8}
\noindent\textbf{Open Problem \thetakeaway:}
\textit{The practical scalability of TEE defenses
for \moai systems remains limited by a deployment-integration
gap, as protecting diverse on-device models requires coordinated
support across model formats, AI frameworks, operator libraries,
delegates, and CPU/GPU/NPU isolation interfaces, yet current
mobile ecosystems still lack widely adopted, developer-transparent
TEE-backed inference stacks.}
\end{takeaway}
\vspace{-1em}

\subsubsection{Model Watermarking}
In contrast to the aforementioned proactive defenses, which aim to prevent unauthorized model access or misuse before and during inference, model watermarking protects on-device models through ownership verification after model extraction, copying, or unauthorized reuse occurs.
Since watermarking for on-device models in \moai systems remains less explored than preventive defenses, we discuss existing watermarking work separately rather than further categorizing it.

Huang et al.~\cite{huang2025themis} proposed THEMIS, the first practical watermarking tool for intellectual property protection of post-deployment on-device models in \moai systems. It lifts the read-only and inference-only constraints of on-device models by reconstructing writable counterparts through Model Rooting and solving watermark parameters through training-free Model Reweighting. Through selective parameter modification, THEMIS enables black-box ownership verification of suspected stolen models while preserving utility on benign inputs.

\begin{observation}
\noindent\textbf{Security-cost Trade-off: }
\textit{Model watermarking improves post-deployment accountability by enabling ownership verification after model extraction or unauthorized reuse, but its cost shifts to utility degradation, offline embedding time, and deployment compatibility. THEMIS~\cite{huang2025themis} achieves over 80\% watermark success rate across evaluated on-device scenarios, but may reduce benign accuracy by up to 12.76\% when label data is missing. In addition, THEMIS successfully watermarks 81.14\% (327/403) of evaluated Google Play AI apps, while failures arise from anti-repackaging mechanisms and unknown model-decryption APIs.}
\end{observation}
\vspace{-1em}

\begin{takeaway}
\refstepcounter{takeaway}\label{op9}
\noindent\textbf{Open Problem \thetakeaway:}
\textit{Although model watermarking enables post-deployment ownership verification for on-device models in \moai systems, its effectiveness remains limited by redeployment-induced watermark fragility, as stolen models may be reused via framework conversion, encryption, or app-level input-output mediation that preserves benign inference while disrupting the trigger responses, confidence patterns, or output semantics used for ownership verification.}
\end{takeaway}
\vspace{-1em}

\subsection{Cross-pillar Analysis of \moai Defenses}
\label{sec:pillar_analysis_defense}
Figure~\ref{fig:pillars} (the lower half) illustrates how the defense categories in \hyperref[sec:defense_taxonomy]{\S\ref*{sec:defense_taxonomy}} and their corresponding open problems (\hyperref[op6]{OP\ref*{op6}}-\hyperref[op9]{OP\ref*{op9}}) map to the three security pillars of \moai systems.
Existing defenses are model-oriented in purpose but lifecycle-dependent in enforcement, with protection realized through pre-release concealment, app-bound authorization, TEE-backed execution, and post-deployment ownership verification.
The mapping further shows that the identified OPs expose the robustness limits of current defenses, since their effectiveness depends on executable equivalence under obfuscation, app-bound authorization in client-side code, TEE-backed inference support across heterogeneous mobile stacks, and watermark verifiability after reuse.

\textbf{Security Pillars to \moai Defenses.}
The defense mapping reveals a model-centric protection orientation in \moai systems, with Pillar II serving as the common locus of protection, Pillar III carrying those protections into runtime-confined execution, and Pillar I mediating the app-level invocation path for authorized model use.
Model obfuscation protects Pillars II and III because it conceals the structures, parameters, and operators of deployed models and makes correct inference behavior available only through authorized runtime execution.
\hyperref[op6]{OP\ref{op6}} captures the residual exposure inherent to this design, as the runtime states materialized during authorized inference provide recovery signals for semantic, structural, or parameter reconstruction.
Model authorization concentrates on Pillar II as it protects model artifacts through conditional correctness, where app-specific credentials and artifact verification govern correct inference behavior.
This authorization check places the trust decision in the deployed mobile stack, so \hyperref[op7]{OP\ref{op7}} concerns the robustness of client-side authorization against reverse engineering, repackaging, or instrumentation.
TEE defenses cover Pillars II and III since they safeguard deployed models through hardware-backed isolation, ranging from full-model execution and security-sensitive partitions to TEE-side recovery of accelerator-offloaded outputs.
The key challenge for such defenses captured by \hyperref[op8]{OP\ref{op8}} lies in making them deployable at scale across heterogeneous mobile inference and isolation stacks.
Model watermarking strengthens Pillar II because it embeds ownership evidence into deployed models and enables behavioral verification of suspected model extraction or reuse.
\hyperref[op9]{OP\ref{op9}} captures the fragility of this evidence after model reuse, as redeployment through model conversion, encryption, or app-level mediation can preserve normal functionality while weakening the behavioral signals needed for verification.

\textbf{Takeaway.}
Although current \moai defenses attach protections to deployed models, their robustness depends on post-release assurance mechanisms that realize those protections through authorized execution, client-side checks, hardware-backed isolation, or post-reuse verification.
These mechanisms are not auxiliary implementations around protected models, but the enforcement basis on which defense guarantees depend after release. 
Their constraints lie in preserving legitimate use of protected models while avoiding recovery handles in authorized execution, bypass targets in client-side checks, portability bottlenecks in hardware-backed isolation, and ownership blind spots in post-reuse verification.
Hence, OP6--OP9 characterize the post-release robustness requirements for model defenses, calling for guarantees that resist execution-based recovery and client-side bypass, remain deployable at scale, and preserve ownership verifiability after reuse in realistic \moai deployments.

\section{Future Directions}
\label{sec:future}


\textbf{Security of On-Device Training in \moai Systems.}
As discussed in \hyperref[sec:attack_taxonomy]{\S\ref*{sec:attack_taxonomy}} and \hyperref[sec:defense_taxonomy]{\S\ref*{sec:defense_taxonomy}}, existing research on \moai security solely focuses on deployed models that are read-only and inference-only, leaving the security implications of on-device training unexplored. 
Emerging on-device training~\cite{ondevice_training} enable models to be updated on end-user devices.
For example, a fashion item recognition \moai system can locally fine-tune its on-device model using user data to better recognize items of interest to users. 
This process exposes model gradients, parameter updates, and user data at runtime, which introduces new security threats beyond those considered in inference-only on-device models.
Hence, future research should explore the expanded threat landscape introduced by on-device training, with particular emphasis on how exposure of the training process enables novel on-device models attacks against \moai systems and motivates the design of corresponding defenses.

\textbf{Security of On-device GenAI in \moai Systems.}
Current research on \moai security predominantly focuses on computer vision tasks, particularly image classification, while security assessment of other domains like natural language processing remains unexplored.
With the rapid evolution of LLMs, on-device GenAI~\cite{ondevice_genai} has become feasible across smartphones to support LLM inference and image generation.
For instance, a conversational \moai system can execute an on-device LLM to process user prompts and generate responses locally, handling sensitive conversational data on end-user devices.
However, on-device GenAI introduces security challenges that differ from those of vision-based models, such as prompt injection, jailbreaking behaviors, and unintended information disclosure.
Therefore, future research should consider on-device security beyond vision-based models and  establish practical threat models that capture the distinctive attack surfaces and defense challenges in on-device GenAI.

\textbf{Security of Agentic \moai Systems.}
Driven by increasingly capable LLMs and emerging mobile tool interfaces, \moai systems are expanding from passive on-device inference toward agentic workflows that connect mobile app contexts, private user data, sensors, OS services, and cross-app interfaces with foundation models executed on device, in the cloud, or through hybrid paths. 
This transition expands the security scope from protecting deployed models and inference pipelines to securing context-to-action chains on end-user devices, where malicious prompts, poisoned app context, compromised tool outputs, or misleading sensor signals may trigger privacy leakage, unauthorized operations, cross-app abuse, or persistent device-state changes without modifying the deployed model artifact.
Thus, future research should move beyond prompt-level defenses and develop end-to-end action governance for agentic \moai systems, with particular emphasis on establishing provenance for mobile context, separating trusted user intent from untrusted environmental content, binding tool and API invocations to task-scoped permissions, enforcing confirmation and rollback for sensitive operations, and auditing agent plans, memory, and actions.

\section{Conclusion}
This paper systematizes the security landscape of \moai systems, where AI models are stored, loaded, and executed on end-user devices. We introduce three security pillars to characterize \moai-specific security properties, develop attack and defense taxonomies across adversarial capabilities and deployment phases, and identify nine open problems that expose how mobile deployment reshape both attack practicality and defense robustness. Looking forward, we outline on-device training, on-device GenAI, and agentic \moai systems, as future directions for extending \moai security landscape. Our SoK can serve as a foundation for principled security analysis and practical defense design for next-generation \moai systems.



\bibliographystyle{plain}
%
\bibliography{reference}

\end{document}